\documentclass[aps,prl,twocolumn,showpacs,superscriptaddress,showkeys]{revtex4-1}

\usepackage{amssymb}
\usepackage{amsmath}
\usepackage{amsfonts}
\usepackage{graphicx}
\usepackage{color}
\usepackage{xspace}
\usepackage{ulem}
\usepackage{mathtools}
\usepackage{hhline}

\usepackage{dcolumn}
\usepackage{bm}
\usepackage{hyperref}
\usepackage{comment}

\usepackage[hyperref,svgnames]{xcolor}

\newcommand{\gsim}{\lower.7ex\hbox{$\;\stackrel{\textstyle>}{\sim}\;$}}
\newcommand{\lsim}{\lower.7ex\hbox{$\;\stackrel{\textstyle<}{\sim}\;$}}

\def\beq{\begin{equation}}
\def\eeq{\end{equation}}
\def\bea{\begin{eqnarray}}
\def\eea{\end{eqnarray}}
\def\bitem{\begin{itemize}}
\def\eitem{\end{itemize}}
\newcommand{\bec}{\begin{center}}
\newcommand{\eec}{\end{center}}
\newcommand{\ba}{\begin{array}}
\newcommand{\ea}{\end{array}}

\def\inv{^{\raise.15ex\hbox{${\scriptscriptstyle -}$}\kern-.05em 1}}
\def\lbar{{\lower.35ex\hbox{$\mathchar'26$}\mkern-10mu\lambda}} 

\let\<=\langle
\let\>=\rangle

\let\+=\uparrow

\def\beqa{\begin{equation}\begin{aligned}}
\def\eeqa{\end{aligned}\end{equation}}

\newcommand{\AddrCoimbra}{Univ Coimbra, Faculdade de Ci\^encias e Tecnologia da Universidade de Coimbra and CFisUC, Rua Larga, 3004-516 Coimbra, Portugal}

\begin{document}

\title{Comparable dark matter and baryon abundances with a heavy dark sector}

\author{Jo\~{a}o G.~Rosa} \email{jgrosa@uc.pt}\affiliation{\AddrCoimbra}
\author{Duarte M.~C.~Silva} 
\affiliation{\AddrCoimbra}

\date{\today}

\begin{abstract}
We propose a scenario that explains the comparable abundances of dark matter (DM) and baryons without any coincidence in the corresponding particle masses. Here, DM corresponds to heavy ``dark baryons" in a hidden MSSM-like dark sector, where the supersymmetry breaking scale can be several orders of magnitude larger than in the visible sector. In both sectors a baryon asymmetry is generated via the Affleck-Dine mechanism, and the smaller dark baryon-to-entropy ratio partially compensates the larger dark baryon masses to give similar densities in the two sectors. The large mass hierarchy also naturally results in an asymmetric reheating of these sequestered sectors. Moreover, this scenario predicts uncorrelated DM and baryon isocurvature perturbations.
\end{abstract}


\maketitle


One of the most puzzling aspects of the mysterious dark matter (DM) component in our Universe is the fact that its abundance is only $\sim 5$ times larger than that of ordinary matter, despite its apparently very different nature. Evidence from galaxy rotation curves, weak gravitational lensing, Cosmic Microwave Background (CMB) anisotropies and large-scale structure of the Universe points towards DM being essentially a cold non-collisional/non-dissipative gas. This coincidence has no clear explanation in standard DM scenarios with e.g.~weakly-interacting thermal relics or axion-like particles. This motivates scenarios with asymmetric DM to explain the comparable DM and baryon abundances  \cite{Nussinov:1985xr, Gelmini:1986zz, Chivukula:1989qb, Barr:1990ca, Kaplan:1991ah, Hooper:2004dc, Kitano:2004sv, Cosme:2005sb, Farrar:2005zd, Suematsu:2005kp, Tytgat:2006wy, Kaplan:2009ag, An:2009vq, March-Russell:2011ang, Cirelli:2011ac, Bai:2013xga, Newstead:2014jva, Ibe:2019yra, Ibe:2021gil, Blinov:2021mdk, Ritter:2022opo}, where a matter-antimatter asymmetry is generated in a dark sector by a mechanism akin to the one producing the observed cosmological baryon asymmetry. While this yields comparable {\it number densities} for baryons and DM, it requires a coincidence between the corresponding particle masses. This could occur e.g.~in the context of mirror world \cite{Hodges:1993yb, Berezhiani:2000gw, Ignatiev:2003js, Berezhiani:2003xm, Berezhiani:2005ek, Berezhiani:2008zza, Ciarcelluti:2010zz, Das:2011gj, Gu:2012fg, Higaki:2013vuv, Gu:2013nya, Lonsdale:2018xwd, Ibe:2019ena, Ritter:2021hgu} or twin Higgs scenarios \cite{Falkowski:2006qq, GarciaGarcia:2015pnn, Reece:2015lch, Farina:2016ndq, Chacko:2018vss}, where the dark sector (partially) mirrors the Standard Model (SM) or its minimal supersymmetric extension (MSSM), potentially restoring an overall left-right symmetry as proposed by Lee and Yang \cite{Lee:1956qn}.

In this Letter, we show that no coincidence between DM and baryon masses is required to explain their similar densities. Our scenario also considers a dark sector that is a copy of the MSSM, with the same gauge group, particle content and gauge coupling at a common grand unification theory (GUT) scale, $M_{GUT}$. There are, however, no (unbroken) symmetries relating the two sectors, which are assumed to be sequestered, i.e.~interact very feebly with each other. Most importantly, the supersymmetry (SUSY) breaking scale may be (very) different in the two sectors, which as we will show affects both the mass of the lightest dark baryons and their number density. However, the dark and visible baryon energy densities are nevertheless comparable, under simple assumptions about the dark quark mass hierarchy.  

A potential motivation for this setup is heterotic superstring theory with the gauge group $E_8\times E_8$, where each $E_8$ factor may be spontaneously broken to $SU(3)\times SU(2)\times U(1)$. If each of these sectors is located in a different region of the six compact extra-dimensions (e.g.~at the bottom of distinct warped throats), they are naturally sequestered from each other and SUSY breaking in the bulk of the extra-dimensions may be communicated in a different way to each sector. Yukawa couplings may also depend on the local extra-dimensional geometry, while ultra-violet (UV) physics like gauge coupling unification generically depends on the bulk geometry and is, hence, common to both sectors. Different SUSY breaking scales may also be due to different couplings between each sector and the hidden sector where SUSY is spontaneously broken, through heavy ($\gtrsim M_{GUT}$) messenger fields. Our analysis is, however, largely independent of the mechanism behind SUSY breaking in both sectors.


We assume that a baryon asymmetry is generated, in both the visible and dark sectors, via the Affleck-Dine (AD) mechanism \cite{Affleck:1984fy}, through the dynamics of scalar flat directions carrying baryon number or $B-L$ \cite{Dine:1995kz, Enqvist:2003gh, Allahverdi:2012ju}. These are an accidental feature of the MSSM (and of its dark copy), since the gauge symmetry leaves several directions in scalar field space along which the potential is constant at the renormalizable level, for unbroken SUSY \cite{Gherghetta:1995dv}. Such directions are, however, lifted by non-renormalizable terms (e.g. from supergravity contributions or couplings to the inflaton superfield) as well as soft scalar mass terms and trilinear A-terms upon spontaneous SUSY breaking. Several MSSM flat directions carry a non-vanishing  $B$ or $B-L$ charge, and the non-renormalizable scalar potential generically includes CP-violating terms, which then allow for the development of a non-vanishing baryonic/$B-L$ charge through the non-equilibrium dynamics of the scalar fields. The potential for a SUSY flat direction $\phi$ is generically of the form:
\begin{eqnarray}\label{potential}
V(\phi)&=&(m_\phi^2-cH^2)|\phi|^2+\left[{\lambda(aH+Am_\phi)\over n M^{n-3}}\phi^n+\mathrm{h.c.}\right]\nonumber\\
&+&\lambda|^2{|\phi|^{2(n-1)}\over M^{2(n-3)}}~.
\end{eqnarray}
Here the terms proportional to the Hubble parameter $H$ are due to $\phi$-inflaton couplings (during inflaton-dominated cosmic eras), while the terms proportional to $m_\phi\sim M_{SUSY}$ are due to the spontaneous SUSY breaking in the hidden sector. The non-renormalizable terms are suppressed by a large mass scale $M$ which we take to be of the order of $M_{GUT}$ or the (reduced) Planck mass $M_P$, and the constants $c, a, A$ and $\lambda$ are typically $\mathcal{O}(1)$ numbers. The order $n$ of the flat direction depends on the number and charges of the superfields it involves, and in the MSSM with unbroken R-parity only $n=4,~6$ flat directions generate baryon number (see e.g.\cite{Enqvist:2003gh}). In supergravity scenarios with minimal K\"ahler terms, one generically has $c<0$, but a non-minimal K\"ahler potential can lead to $c>0$ so that the scalar field develops a large expectation value during inflation, when $H\gg m_\phi$:
\begin{eqnarray} \label{minimum}
|\phi|\sim \left(\sigma H M^{n-3}/ \lambda\right)^{1\over n-2}~,
\end{eqnarray}
where $\sigma\sim \mathcal{O}(1)$ is a combination of $c, a$ and $n$. Note that the field is critically damped during inflation for $|c|\sim \mathcal{O}(1)$ and that its complex phase at this stage is set by the phase of the $a$ parameter. Since, as we will see below, we are interested in scenarios with low reheating temperatures, inflation is generically followed by a parametrically long inflaton matter-dominated era, when the latter behaves as non-relativistic matter before decaying and reheating the Universe. The flat direction remains critically damped in this phase, closely following the evolving minimum (\ref{minimum}) until $H\sim m_\phi$. Once $H\lesssim m_\phi$ the flat direction becomes underdamped and starts oscillating about the new minimum at the origin. However, the generic mismatch of the complex phases of the $a$- and $A$-terms endows the field with a `phase velocity' $\dot\theta$, thus generating a baryon number density $n_B= 2\beta|\phi_0|^2\dot\theta$, where $|\phi_0|$ corresponds to the value of the potential minimum (\ref{minimum}) at the onset of oscillations, $H\sim m_\phi$, and $\beta$ is its $B/B-L$ charge. This typically yields $n_B\lesssim n_\phi$, i.e.~the AD field develops an $\mathcal{O}(1)$ baryon asymmetry from this stage onwards. Since the CP-violating $A$-term decreases as $\phi$ oscillations are damped by expansion, baryon number (or $B-L$) is approximately conserved at late times.

Both the inflaton and the AD field $\phi$ behave as non-relativistic matter until the former decays, and so the ratio between their energy densities remains constant until reheating, which allows one to easily compute the resulting baryon-to-entropy ratio: 
\begin{eqnarray} \label{eta}
\eta={n_B\over s}\approx \beta {n-2\over 6(n-3)}\left({M^{n-3}\over \lambda M_P^{n-2}}\right)^{\delta_n}T_R M_{SUSY}^{\delta_n-1}~,
\end{eqnarray}
where $\delta_n = 2/(n-2)$ and $T_R$ is the reheating temperature. In the above expression we have omitted $\mathcal{O}(1)$ factors and taken $m_\phi\sim M_{SUSY}$. This ratio becomes constant after reheating, assuming there are no additional sources of baryon number or entropy, with the AD field eventually decaying in a $B$/$B-L$-conserving fashion \cite{foot1} and transferring the produced asymmetry into quarks that subsequently hadronize. Also note that electroweak sphalerons \cite{Klinkhamer:1984di} partially convert an initial lepton asymmetry into a baryon asymmetry, while preserving $B-L$. In the visible sector, one then obtains the observed value $\eta_B\sim 10^{-10}$ for e.g.~$\lambda=1$, $M=M_P$, $M_{SUSY}\sim 1$ TeV and $T_R\sim 10^9$ GeV ($n=4$) or $T_R\sim 100$ GeV ($n=6$). 

Note that, for $n=6$, thermal effects \cite{Allahverdi:2000zd, Anisimov:2000wx, Anisimov:2001dp} play a negligible role in the AD dynamics, since, before reheating $T^4\sim T_R^2 H M_P$ (for constant inflaton decay width), and hence $\phi/T\sim (M^3/T_R^2M_P)^{1/4}\gg 1$. Fields coupled to the flat direction are thus typically much heavier than the ambient temperature, and even integrating out their effects induces only a residual thermal mass $ m_{\phi,T}^2\sim T^4/\phi^2 \ll T_R^2$ (see e.g.~\cite{Allahverdi:2012ju, Anisimov:2000wx}).

Both the visible and dark sector asymmetries are given by an expression of the form \eqref{eta}, with strong interactions in both sectors ensuring the efficient annihilation of any symmetric component. While there could be multiple flat directions in both sectors with $B/B-L\neq 0$ and $c>0$ \cite{foot2}, those of order $n=6$ acquire the largest field values and hence yield the dominant contribution to the baryon asymmetry in both sectors. The mass scale $M$ should also be common to both sectors (which share the same UV physics), so that up to $\mathcal{O}(1)$ factors we find $\eta_D/\eta_B \sim (M_{SUSY}^{D}/M_{SUSY})^{\delta_6-1}$, with $\delta_6=1/2$, where $M_{SUSY}^D$ denotes the dark sector's SUSY breaking scale. This implies, in particular, $\eta_D<\eta_B$ if the SUSY breaking scale is larger in the dark sector. We note, however, that in the particular case where only $n=4$ flat directions have $c>0$, $\eta$ is independent of $M_{SUSY}$. 

 The distinct SUSY breaking scales in both sectors also make the corresponding light baryon masses different, assuming that, as for protons and neutrons, they are mostly due to (dark) QCD binding energy, which is the case for sufficiently light dark up and down quarks. Thus, dark baryon masses are set by the dark QCD scale at which $\alpha_s(\Lambda_{QCD})\sim 1$ \cite{foot3}. Since superpartners decouple from the gauge coupling running at energies below their mass threshold, $\Lambda_{QCD}^D$ intrinsically depends on $M_{SUSY}^{D}$. 

As we are interested in finding the ratio of baryon and dark baryon masses up to $\mathcal{O}(1)$ factors, it is sufficient to consider the 1-loop running of the strong coupling constant (in the $\overline{\mathrm{MS}}$ scheme) to estimate $\Lambda_{QCD}^D$, with $d\alpha_s/d\log Q=-b\alpha_s^2/2\pi$ in both sectors. Since superpartners generically have similar masses, we take $b_{MSSM}=3$ for $Q>M_{SUSY}^D$, with all superpartners decoupling below this scale \cite{foot4}. It is also natural to assume an $\mathcal{O}(1)$ dark top Yukawa coupling, as in the MSSM, and thus that the dark electroweak scale is set by $M_{SUSY}^D$, so that we expect the ratio $m_t/M_{SUSY} $ to be comparable in the visible and dark sectors. We will focus on the regime where all other dark quarks are lighter than $\Lambda_{QCD}^D$, with Fig.~1 illustrating the running of the dark $\alpha_s$ setting the UV  boundary condition $\alpha_s^{-1}\simeq 25.6$ at $M_{GUT}\simeq 2\times 10^{16}\ \mathrm{GeV}$ in both sectors. 
\begin{figure}[h]
\centering\includegraphics[scale=0.95]{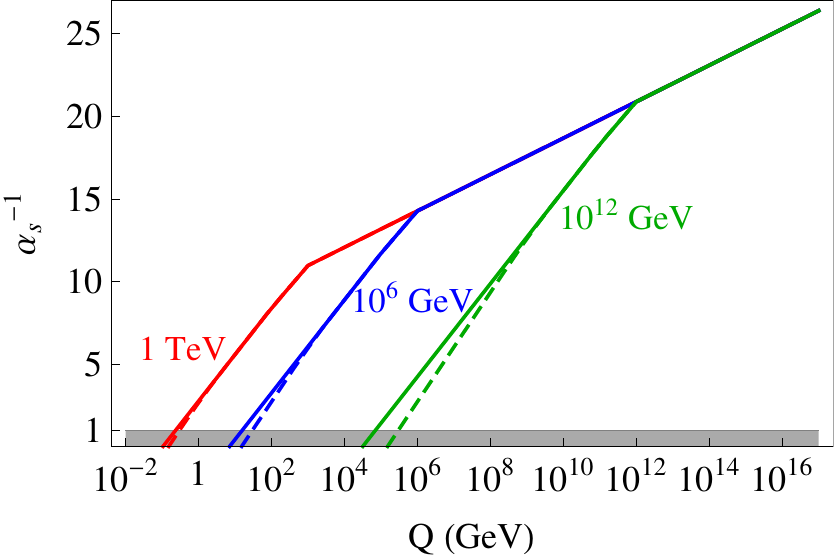}
\caption{Running strong coupling constant in the dark sector for different values of the SUSY breaking scale $M_{SUSY}^D$ as labeled, assuming common boundary conditions at the GUT scale and considering that only the dark top quark is heavier than $\Lambda_{QCD}^D$ (solid lines) or that the dark top, bottom and charm quarks have the same mass ratios as their visible counterparts (dashed lines). The gray region corresponds to the non-perturbative regime.}
\end{figure}

We then find for both sectors:
\begin{eqnarray} \label{eta}
\Lambda_{QCD} = M_{SUSY}^{1-\epsilon} M_{GUT}^\epsilon e^{-{2\pi\over b_5}(\alpha_{GUT}^{-1}-1)}\left({m_t\over M_{SUSY}}\right)^{-b_6/b_5}
\end{eqnarray}
where $b_n=(33-2n)/3$ is the QCD beta-function coefficient for $n$ quark flavours and $\epsilon=b_{MSSM}/b_6\simeq 0.39$, yielding $\Lambda_{QCD}\sim 200(M_{SUSY}/1\ \mathrm{TeV})^{0.6}$ MeV for $m_t/M_{SUSY}\simeq 0.1$. Including dark bottom and charm quarks heavier than $\Lambda_{QCD}^D$ in the running of $\alpha_s$ (decoupling below their respective mass thresholds) may slightly change this result, as illustrated in Fig.~1 for the case of identical quark mass ratios in the visible and dark sectors. However, if they are not much heavier than $\Lambda_{QCD}^D$ we may neglect these effects since $\alpha_s$ runs logarithmically.

Hence, the opposite dependence on the SUSY-breaking scale of $\eta$ and  $\Lambda_{QCD}$ yield:
\begin{eqnarray} \label{eta}
{\rho_{DM}\over \rho_B}\!=\! {m_{D}\eta_D\over m_B\eta_B}\!=\! {\Lambda_{QCD}^D\eta_D\over \Lambda_{QCD}\eta_B}\!=\! \left({M_{SUSY}^D\over M_{SUSY}}\right)^{\delta_n-\epsilon}\!\!\!\!\!\!\!\times \mathcal{O}(1)\,,
\end{eqnarray}
where the $\mathcal{O}(1)$ factors reflect the potential differences between the AD potential parameters, quark mass hierarchies and superpartner mass splittings in the two sectors. Most strikingly, the DM/baryon abundance ratio is only mildly dependent on the SUSY breaking scale ratios, with $\delta_6-\epsilon\simeq 0.1$ for $n=6$ flat directions producing the dark and visible baryon asymmetries. This implies that DM and baryons may have comparable abundances even if dark baryons are much heavier than protons or neutrons, which is the main result of this work, and is illustrated in Fig.~2.

\begin{figure}[htbp]
\centering\includegraphics[scale=0.95]{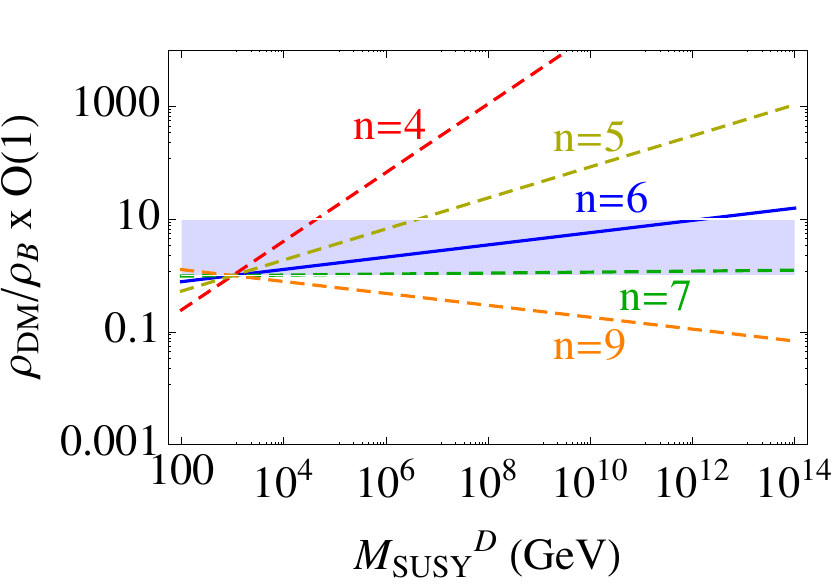}
\caption{Ratio between dark and baryonic matter energy densities (up to $\mathcal{O}(1)$ factors) as a function of the dark SUSY breaking scale, taking $M_{SUSY}=1$ TeV in the visible sector, for flat directions lifted at different order $n$ as labeled.}
\end{figure}
As one can see in this figure, for $n=6$ baryonic and dark AD fiat directions (BAD/DAD, respectively) the DM and baryon abundances differ by less than an order of magnitude for dark SUSY breaking scales up to $\sim 10^{12}$ GeV, i.e.~for dark baryons with masses up to $\sim 100$ TeV. We also show in this figure the results for flat directions of different order (but the same in both sectors), where one can see that $n=4$ flat directions can only accommodate comparable abundances if the SUSY breaking scales in both sectors differ by at most 2 orders of magnitude. Flat directions with odd $n$ also yield $\rho_{DM}\sim\rho_B$ for a wide range of masses, namely $n=7$ for which $\delta_7-\epsilon\simeq 0.01$, but as mentioned above these only contribute to $\eta_{B/D}$ if R-parity is broken at early times. 

The more promising $n=6$ scenario is also appealing since DM is naturally more abundant and heavier than baryons. A larger dark SUSY breaking scale implies also that dark squarks, sleptons and Higgs fields are heavier than their corresponding visible counterparts. Since the inflaton field, $\Phi$, is expected to be a gauge singlet, it cannot decay directly into chiral fermions through renormalizable operators (except potentially right-handed neutrinos, which need not be present in our setup), coupling only to scalar degrees of freedom. Hence, while the inflaton may decay directly into visible sector scalars, this may be kinematically forbidden for the much heavier dark scalars, which can only mediate inflaton decays into lighter dark particles as virtual modes. The inflaton partial decay width into the dark sector may thus be suppressed by powers of $m_\Phi/M_{SUSY}^D\ll 1$ with respect to the visible sector, thus resulting in an asymmetric reheating of the Universe that populates mostly the MSSM sector.

Moreover, since the two sectors are assumed to be sequestered from each other, no thermalization between them should occur and most of the entropy in the Universe corresponds to SM degrees of freedom. In particular, light particles in the dark sector, such as the dark photon or even dark neutrinos (depending on how their mass is generated), will not contribute significantly to the number of relativistic species at the nucleosynthesis and recombination epochs. We note, however, that current limits from Planck still allow for a dark photon and up to 3 dark neutrinos in equilibrium with the SM thermal bath until the (visible) electroweak phase transition \cite{Planck:2018vyg}, so we need not be too strict on precluding the thermalization of the dark and visible sectors.

Interestingly, the dark neutron may be lighter than the dark proton, yielding an essentially collisionless and non-dissipative DM candidate that could explain, in particular, why it does not form disks on galactic scales. In fact, in the SM the positive isospin quark is only lighter than its negative isospin counterpart in the first quark family, a peculiar feature that need not be shared with the dark sector. In fact, if the dark proton-neutron mass difference is large enough, dark nuclei may be unstable. Within our framework, we may even envisage scenarios with multiple MSSM-like sectors with comparable baryon energy densities, but where only a single ``visible" sector has a stable light proton and hence a non-trivial nuclear physics and chemistry, with all others making up the DM.

A key requirement is that the lightest baryon is sufficiently stable in both sectors. While this is conventionally ensured in SUSY models by R-parity, this would imply a stable lightest superpartner (LSP) contributing to the DM density, potentially spoiling our conclusions. There are, however, broken R-parity models where proton stability is ensured by other symmetries (e.g. \cite{Tatar:2006if}) that allow the LSP to decay, and which in the present setup could be applied to both visible and dark sectors.

The assumed sequestering of the visible and dark sectors makes it hard to directly test our hypothesis, given also the potentially much larger masses of dark sector particles. We may nevertheless envisage small Higgs-portal couplings or kinetic mixing between the dark and visible photons that could allow for laboratory or astrophysical signatures, although this depends on the particular implementation of the proposed scenario that have no significant impact on addressing the DM-baryon coincidence problem, which is our goal in this Letter.

There may, however, be a smoking-gun for AD `mattergenesis' in our scenario if the phases of the complex BAD and DAD fields are light during inflation, $m_\theta\lesssim H_I$ (where $H_I$ denotes the Hubble parameter during inflation), which is the case for $a\lesssim 1$ in the potential \eqref{potential}. In this case the phase of the BAD/DAD fields is essentially frozen at their initial value in the inflationary patch leading to the presently observable Universe (the so-callled {\it spontaneous (dark) baryogenesis} scenario), but exhibit quantum fluctuations that are stretched by inflation to sizes beyond the Hubble radius. This results in super-horizon fluctuations in $\eta_B$ and $\eta_D$ that lead to isocurvature perturbations in the CMB anisotropy spectrum. Most importantly, our model predicts {\it uncorrelated }baryon and DM isocurvature modes, since these arise from independent AD fields. Following e.g.~\cite{Enqvist:2003gh}, this yields a contribution from DM/baryon modes to the total isocurvature power spectrum:
\begin{eqnarray} \label{isocurvature}
\Delta_{B/DM}^2\sim \left({\Omega_{B/DM}\over \Omega_m}\right)^2\left({\lambda\over \sigma}\right)^{1/2}\!\!\left({H_I\over M}\right)^{3/2}\!\!e^{-{2\over 3}{m_\theta^2\over H_I^2}N_e}
\end{eqnarray}
up to $\mathcal{O}(1)$ factors, where $\Omega_{B/DM}$ denotes the present baryon/DM abundance, $\Omega_m=\Omega_B+\Omega_{DM}$ and $N_e\simeq 55$ is the number of e-folds of inflation after CMB scales become larger than $H_I^{-1}$. For $H_I\simeq 10^{13}$ GeV and $M=M_{GUT}-M_P$, current CMB bounds on matter isocurvature modes $\Delta_m^2\lesssim 7.2\times 10^{-11}$ from the Planck satellite \cite{Planck:2018jri} already impose $m_\theta/H_I\gtrsim 0.2-0.6$ (assuming equal masses for the BAD and DAD phase fields). Planned CMB missions may be able to improve these bounds and test our model, but cannot distinguish between baryon and DM isocurvature perturbations. However, future Hydrogen-mappings using the 21 cm line should be sensitive to the baryon isocurvature perturbations only \cite{Kawasaki:2011ze}. In conjunction with CMB measurements, there is thus a realistic possibility of testing the mattergenesis scenario proposed in this Letter, although we note that isocurvature perturbations are exponentially suppressed for $m_\theta\sim H_I$, in which case the above-mentioned direct detection avenues for the dark sector need to be explored.

In summary, we are proposing a simple and fairly generic model where dark baryons in a copy of the MSSM account for DM with an abundance naturally comparable to ordinary baryons despite being much heavier due to the different SUSY breaking scale. Our main conclusions rely on a common UV completion of both MSSM copies at the GUT scale and the existence of light dark  up and down quarks, with some mild model-dependence on the mass hierarchy of the heavy dark quarks. Although we have illustrated our results for TeV-scale superpartners in the visible sector, which so far have not been detected at the LHC, our natural solution to the DM/baryon coincidence problem relies on the ratio between the SUSY breaking scales in both sectors rather than their actual values. We nevertheless hope that our analysis motivates further exploration of more realistic SUSY scenarios with the main ingredients outlined in this work.

\vspace{0.5cm}

{\bf Acknowledgments:} This work was supported by the CFisUC project No.~UID/FIS/04564/2020 and by the FCT-CERN grant No.~CERN/FIS-PAR/0027/2021. 

\vfill


\begin{thebibliography}{99}
\bibitem{foot1}
The AD field may first break into Q-ball solitonic configurations before decaying, although this does not affect our discussion.
\bibitem{foot2}
We assume that there is no significant dynamical interplay between different flat directions.
\bibitem{foot3}
Although there is no rigorous proof of the relation between non-perturbative effects and quark confinement in QCD, these two aspects are widely believed to be interconnected.
\bibitem{foot4}
We are neglecting the effects of threshold corrections, which would have to be included for a precise determination of the confinement scale (see e.g. \cite{Ritter:2022opo}), which is not our goal in this Letter.
\end{thebibliography}
\end{document}